\documentclass{iopart}
\usepackage{color, colortbl}

\definecolor{blue}{cmyk}{1,1,0.25,0}
\definecolor{green}{cmyk}{0.64,0,0.95,0.40}
\definecolor{red}{cmyk}{0,1,1,0}
\definecolor{pink}{cmyk}{0,0.82,0,0}
\definecolor{orange}{cmyk}{0,0.61,0.87,0}

\usepackage{hyperref}
\usepackage{pslatex}

\usepackage[dvips]{graphicx}

\usepackage{enumerate}

\usepackage{array}
\usepackage{multirow}

\begin{document}
%\draft                       % this command makes pacs numbers print

%-------------------------------------------------------------------

\title[Influence on the cooling-rate on the physical properties of glassy GeS$_2$]{Influence of the cooling-rate on the glass transition temperature and 
the structural properties of glassy GeS$_2$: an {\em ab initio} molecular dynamics study}
\author{S\'ebastien Le Roux and Philippe Jund}
\address{Laboratoire de Physicochimie de la Mati\`ere Condens\'ee - Institut Charles Gerhardt\\ 
Universit\'e Montpellier 2, Place E. Bataillon, Case 03,\\
34095 Montpellier, France\\
}

%------------------------------------------------------------------------

\begin{abstract}

Using density-functional molecular dynamics simulations we analyzed the cooling-rate effects on the 
physical properties of GeS$_2$ chalcogenide glasses. Liquid samples were cooled linearly in time 
according to $T(t) = T_0 - \gamma t$ where $\gamma$ is the cooling rate. We found that our model leads
 to a promising description of the glass transition temperature $T_g$ as a function of $\gamma$ 
and gives a correct $T_g$ for experimental cooling rates. We also investigated the dependence of the structural properties on the cooling rate. We show that, globally, the properties determined from our simulations 
are in good agreement with experimental values and this even for the highest cooling rates. In particular,
our results confirm that, in the range of cooling rates studied here, homopolar bonds and extended charged regions
are always present in the glassy phase. Nevertheless in order to reproduce the experimental intermediate 
range order of the glass, a maximum cooling rate should not be exceeded in numerical simulations.
\end{abstract}

%-------------------------------------------------------------------------
\maketitle
%-------------------------------------------------------------------------

\section{Introduction}

During the last fifteen years the unceasingly growing interest for chalcogenide glasses has lead to numerous works. This is in particular true for germanium disulfide glasses for which a large amount of experimental \cite{1999JPCS...60.1473H,1986PhRvB..33.5421B,2001JNCS...279.186,1996JNCS...202.248} and theoretical \cite{ 2003PhRvB..67i4204B, 2004PhRvB..70r4210B, 2004PhRvB..69f4201B, PhysRevB.60.R14985} studies have been carried out. 
Indeed its known properties used for example in optical amplifiers, memory switching devices or anti-reflection coatings \cite{1999JNS...243.116} make him a good candidate for intensive research.\\
Among the different research means, molecular dynamics (MD) simulations can be a very interesting tool to provide detailed 
information on the physical properties of such glassy systems. Firstly because they allow to investigate the structure in full 
microscopic detail giving access to the position of the atoms and, secondly because they are useful to study dynamical phenomena 
accessible to such simulations {\it e.g.}, for time scales between 10$^{-13}$ and 10$^{-8}$ s.\\
If the temperature of a liquid is decreased so much that the relaxation time of the system becomes longer than the time 
scale of the computer simulation or of the experiment, the system undergoes a kinetic arrest and, provided that it does not crystallize, will undergo a glass transition and remain trapped in a disordered configuration.
However it has been demonstrated in both, experiment \cite{1992PhRvB..4611318B, 1994PhRvB..49.3124B} and computer simulations \cite{1989PhRvA..40.6007M, 1996PhRvB..5415808V}, that the properties of the resulting glass, like the density or the glass transition temperature, will depend on its thermal history and in particular on the rate at which the sample is cooled down.\\
Previous studies \cite{2003PhRvB..67i4204B, 2004PhRvB..70r4210B, 2004PhRvB..69f4201B} have already validated our "Cook and Quench" model 
to produce and study GeS$_2$ glasses using approximate {\it ab initio} molecular dynamics simulations. Nevertheless, because of the time 
scale of our computer simulations which is many orders of magnitude shorter than the typical experimental one, the glass transition temperature, for example, appears to be significantly higher than the one observed in the laboratory ($\sim$750K \cite{1986PhRvB..33.5421B, 2001JNS...293.169, PhysRevLett.78.4422}) and thus it is necessary to see how the 
properties of the so-obtained glass depend on the way it was produced.\\
Thus, in the present paper, we focus on the cooling rate effects on some physical properties of glassy GeS$_2$.
Firstly we investigate how it does affect the glass transition temperature $T_g$ and secondly  we study how it does affect the 
structural properties of the glassy samples.\\
The paper is consequently organized as follows. In section \ref{theo} we briefly present the theoretical model used in our calculations, results and discussions are presented in section \ref{res}, and finally in section \ref{conclu} we summarize the major conclusions of our work.

\section{\label{theo}Theoretical framework}

Computations were performed using Fireball96, an approximate {\it ab initio} molecular dynamics code based on the local-orbital electronic structure method developed by Sankey and Niklewski \cite{1989PhRvB..40.3979S}. 
The electronic structure is described using density functional theory (DFT) \cite{1964PhRv..136..864H} within the local density approximation (LDA) \cite{1965PhRv..140.1133K} and the non-local pseudo-potential scheme of Bachelet, Hamann and Schl\"uter\cite{1982PhRvB..26.4199B}. 
To reduce the CPU time we used the non-selfconsistent Harris functional \cite{1985PhRvB..31.1770H} with a set of four atomic orbitals (1 ``s'' and 3 ``p'') per atom that vanish outside a cutoff radius of 5a$_0$ (2.645\AA).
This model has been successfully used the last ten years for several different chalcogenide systems \cite{2003PhRvB..67i4204B,2004PhRvB..70r4210B,2004PhRvB..69f4201B,2001PhRvB..64j4206L,1997PhRvB..56.3054C,2000JPCM...12L..21O}.\\
All the calculations of the present simulations were performed in the microcanonical ensemble, with
a time step $\Delta$t=2.5fs and using only the $\Gamma$ point to sample the Brillouin zone. The initial configuration of our system was a crystal of $\alpha$-GeS$_2$ in a cubic cell of 19.21\AA\ containing 258 atoms with standard periodic boundary conditions, melted at 2000 K for 60ps in order to obtain an equilibrium liquid. Then, we relaxed this system further at 2000K for 50ps and we choose five different liquid samples (approximately every 10ps) during 
this process. Each sample was then quenched down to 300K through the glass transition temperature $T_g$. Quenches were carried out by a linear velocity rescaling according to $T(t) = T_0 - \gamma t$ where $\gamma$ is the cooling rate. Six different cooling rates were used: 3, 0.6, 0.3, 0.12, 0.09 and 0.06$\times$10$^{14}$ K/s.
At 300K each sample was relaxed over 50ps, {\it i.e.}, 20000 time steps. 
Configurations were saved every 20 steps and the results were averaged for each sample over these 1000 configurations. Furthermore we averaged the results for the 5 samples of each cooling rate thus all the data presented below have been averaged over 5$\times$1000 configurations, and the error bars (when given) represent the usual standard deviation. 

\section{\label{res}Results}

\subsection{Glass transition}

\begin{figure}[h!]
\begin{center}
\includegraphics*[width=11cm, keepaspectratio=true, draft=false]{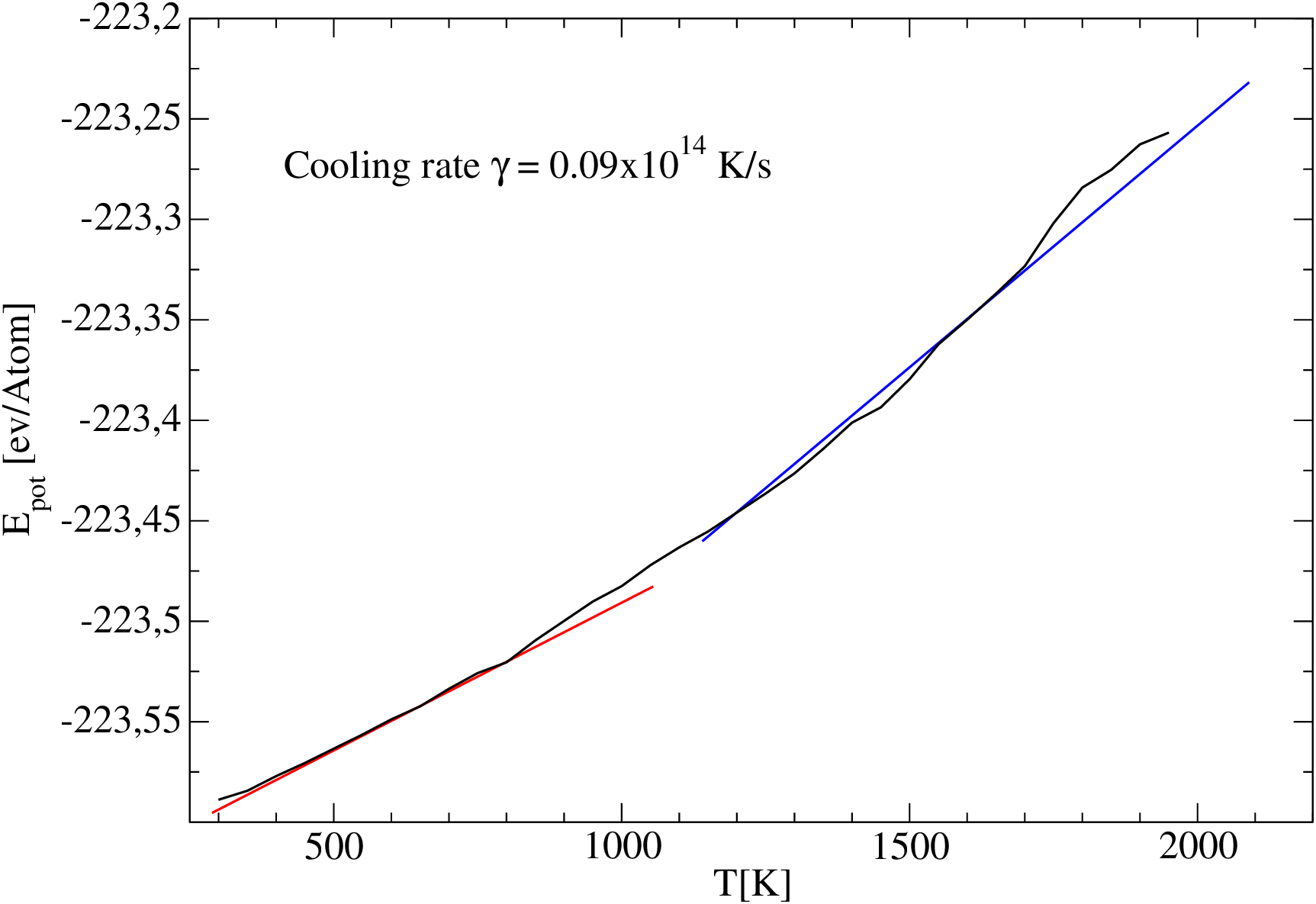}
\end{center}
\caption{\label{ftg} Average potential energy as a function of the temperature and linear regressions above and below the glass transition temperature for a sample cooled at 0.09$\times$10$^{14}$ K/s.}
\end{figure}

\begin{figure}[h!]
\begin{center}
\includegraphics*[width=11cm, keepaspectratio=true, draft=false]{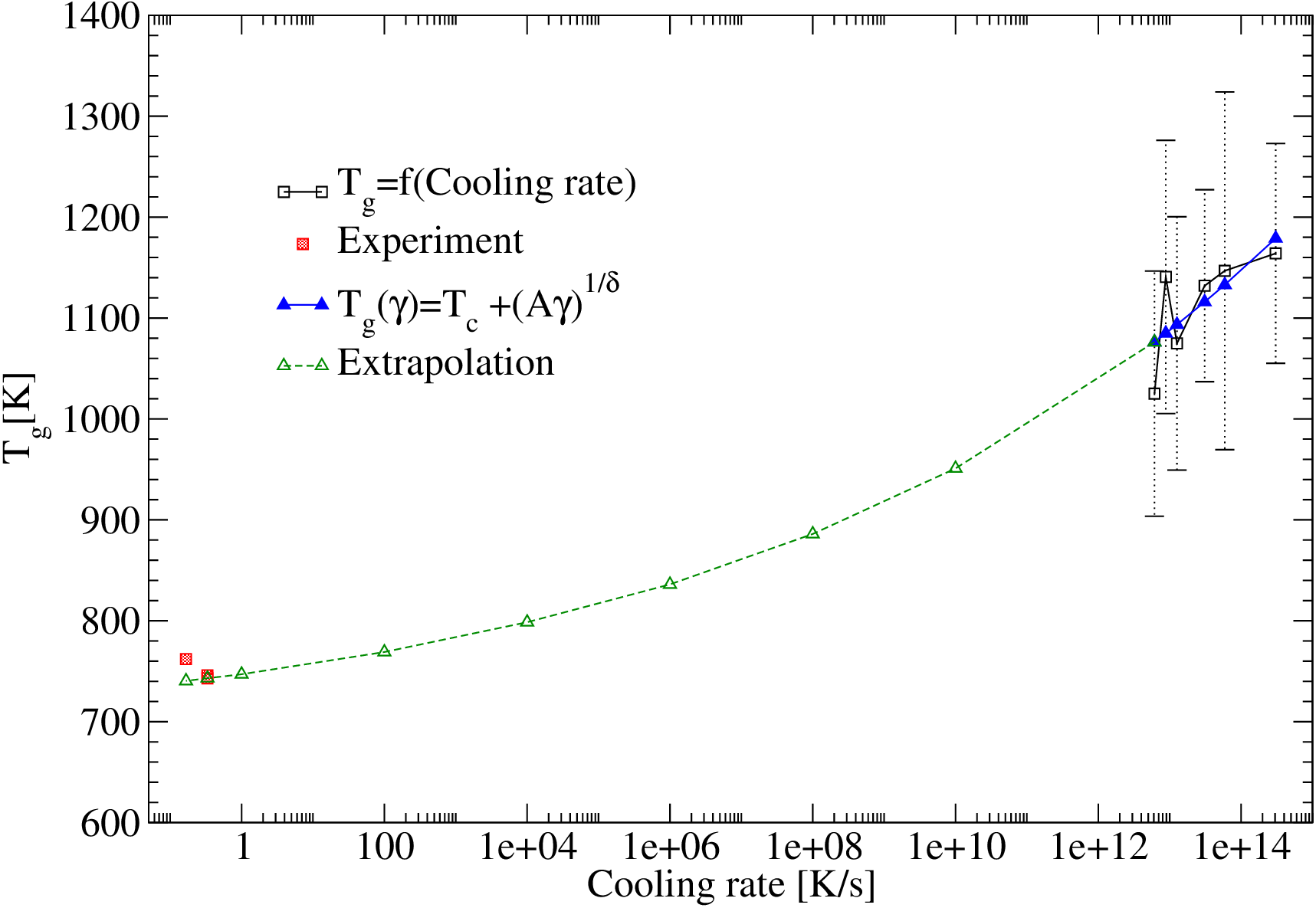}
\end{center}
\caption{\label{Tg} Glass transition temperature as a function of the cooling rate, fit and extrapolation to very low cooling rates - comparison with experiment \cite{1986PhRvB..33.5421B,2001JNS...293.169,PhysRevLett.78.4422}.}
\end{figure}

As usually done in simulations we tried to identify the glass transition of our samples by representing the evolution of the average potential 
energy as a function of the temperature during the quench (an example is proposed for one of the samples cooled at 0.09$\times$10$^{14}$ K/s (Fig. \ref{ftg})), the glass transition being localized by the change in the evolution of the energy.
The glass transition temperature $T_g$ is determined by the intersection of the two linear regressions at high and low temperature (Fig. \ref{ftg}). 
For each cooling rate we averaged the values of the $T_g$ obtained for each of the five GeS$_2$ samples, thus we obtained an evolution of the average $T_g$ with the cooling rate. As shown in Fig. \ref{Tg} the error bars remain huge indicating an insufficient sampling (it is
worth noting that the computer time needed to perform the simulations for the lowest values of $\gamma$ is of the order of
12 weeks for one sample) and it is therefore difficult to extract an accurate description of the evolution of $T_g$ with 
the cooling rate. Nevertheless we attempted to fit these average simulation values assuming a power-law dependence of $T_g$ with the cooling rate 
as suggested by the mode-coupling theory \cite{1992JRPP...55.241} :
\begin{equation}
T_g(\gamma)= T_c + (A\gamma)^{1/\delta}
\end{equation}
with $T_c=676\pm75 K$, $A=3.0\pm2.3\times 10^{31}$ and $\delta =17.1\pm0.1$ (the errors given for the parameters of the fit have been evaluated by fitting the extreme (lowest and highest) Tg values enclosed in the error bars).
We observed that a variation of $\gamma$ by about 1 decade gives rise to a variation of $T_g$ of about 75K which is not that larger, compared to the difference of magnitude of the cooling rates, than the variation of 10K measured in real experiments for different materials \cite{1992PhRvB..4611318B,1994PhRvB..49.3124B,1992JRPP...55.241}.
Then by extrapolating the results of our fit down to usual experimental cooling rates i.e. 10$^0$-10$^5$ K/s, we observed at this scale that a variation of $\gamma$ by about 1 decade gives rise to a variation of $T_g$ of about 13K, which is in agreement with the experimental variation of 10K previously mentioned.
We have represented in Fig. \ref{Tg} the result of this work together with a few experimental data \cite{1986PhRvB..33.5421B,2001JNS...293.169,PhysRevLett.78.4422}. The agreement between the extrapolation of the fit and the experimental values of $T_g$ shows that our 
model is able to give a correct {\em tendency} of the variation of $T_g$ with $\gamma$, even though the poor statistics prevents us
from having accurate estimates of $T_g$ at a specific (high) cooling rate. This can be improved with more simulations in order
to reduce the error bars.
% but the extrapolation to realistic cooling rates (which is the measure of the quality of the fit) would probably not be affected.

\subsection{Structural properties}

\begin{figure}
\begin{center}
\includegraphics*[width=17cm, keepaspectratio=true, draft=false]{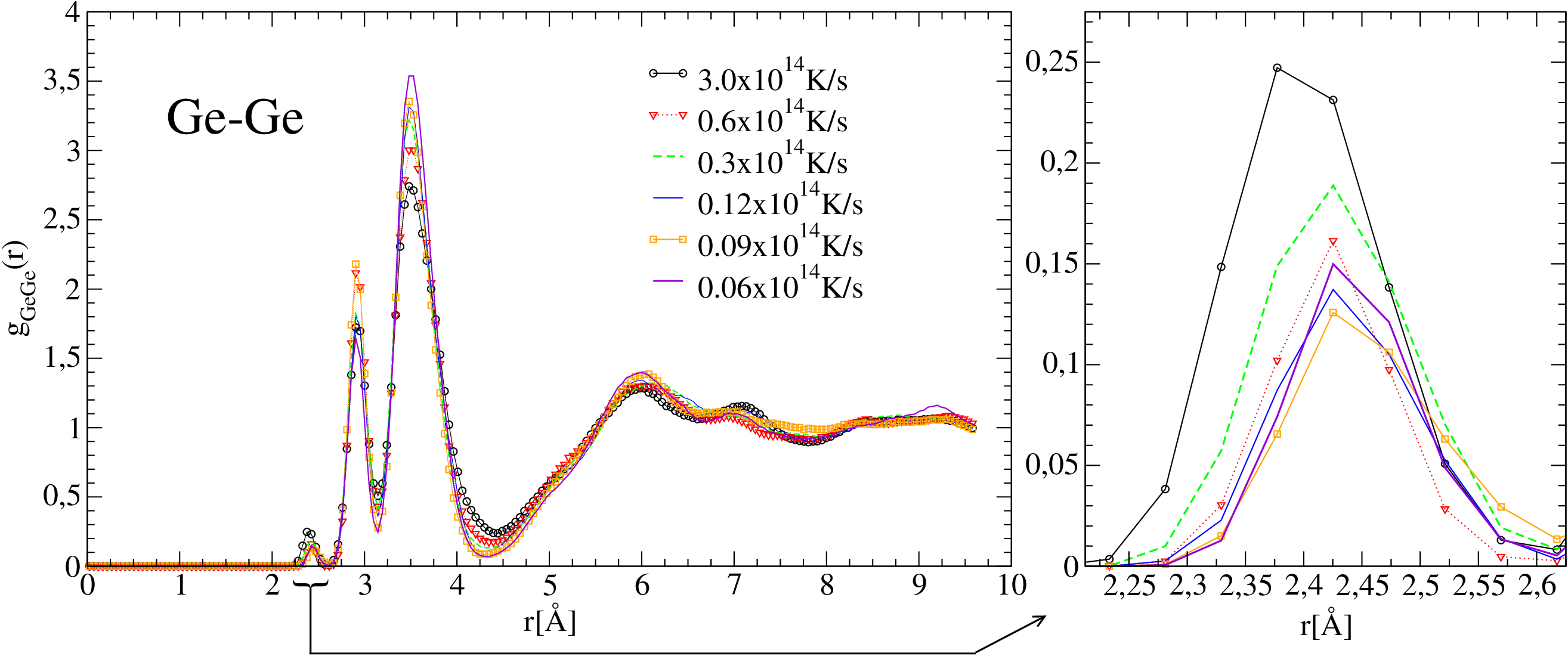}
\end{center}
\caption{\label{GrGeGe} \it Partial radial pair distribution function g(r)[Ge-Ge] as a function of the cooling rate. Enlargement corresponds to details of the peak due to homopolar bonds.}
\end{figure}

\begin{figure}
\begin{center}
\includegraphics*[width=17cm, keepaspectratio=true, draft=false]{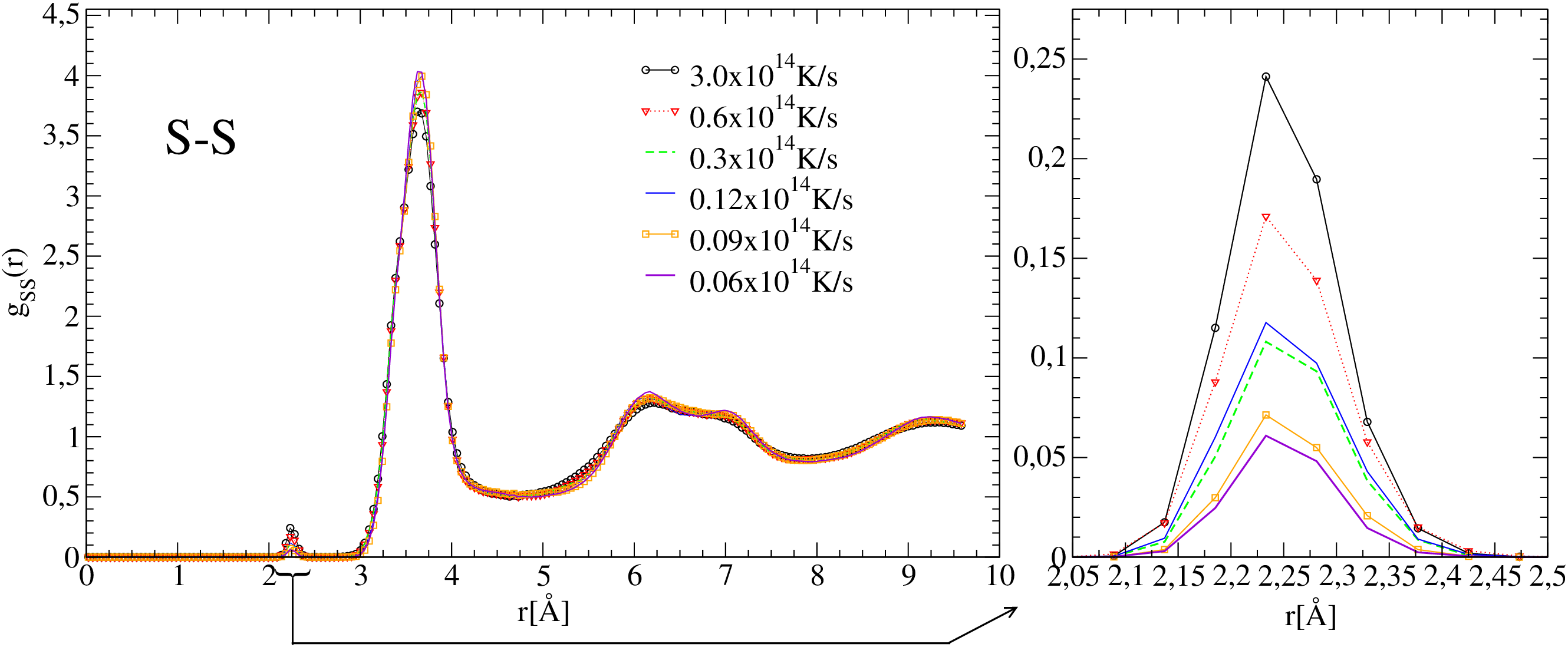}
\end{center}
\caption{\label{GrSS} \it Partial radial pair distribution function g(r)[S-S] as a function of the cooling rate. Enlargement corresponds to details of the peak due to homopolar bonds.}
\end{figure}

\subsubsection{Radial pair correlation functions and bonding properties}

\noindent\\\\In glassy GeS$_2$ the basic building blocks are GeS$_4$ tetrahedra, connected together forming a random network.
The structural disorder is reflected by the absence of long range order and by the wide distribution of bond lengths and bond angles.
Structural information may be extracted from the radial pair correlation function $g(r)$ which can be defined for a given $\alpha,\beta$ pair by:
\begin{equation}
\label{grij}
{\tt{g}}_{\alpha \beta}(r)= \frac{V}{4\pi r^{2} \rho N c_{\alpha} c_{\beta} } \sum_{i \ne j} \delta (r-r_{ij})
\end{equation}
where $\rho$ is the number density of the system, $c_{\alpha}$ the fraction of species $\alpha$ in the system, $i$ the atoms of species $\alpha$ and $j$ the atoms of species $\beta$.
For each cooling rate we averaged the radial pair correlation functions g$_{\alpha \beta}$(r) of the five samples, so we are able to compare the evolution of the average g$_{\alpha \beta}$(r) according to the cooling rate (Fig. \ref{GrGeGe} and \ref{GrSS}).\\
The bond lengths appear to be in good agreement with experimental data\cite{1986PhRvB..33.5421B} since we find 2.23\AA\ for the Ge-S bond (expe.: 2.21\AA), 2.91\AA\ and 3.49\AA\ for respectively the edge and corner sharing Ge-Ge connections (expe: 2.91\AA\ and 3.42\AA). This good agreement is true even for the highest cooling rates. 
The main influence of the cooling rate is reflected in the small peak corresponding to homopolar bonds between 2.2-2.6\AA\ for Ge (Fig. \ref{GrGeGe}) and 2.1-2.45\AA\ for S (Fig. \ref{GrSS}). Our results indicate that the number of homopolar bonds decreases with the 
cooling rate and it is therefore justified to address the question of the existence of homopolar bonds at experimental cooling rates
which is still an open question (Cai and Boolchand using Raman scattering experiment, found the existence of homopolar bonds in glassy 
GeS$_2$ \cite{PhilosMagB.82.1649} while Petri and Salmon found no evidence of such bonds in $g$GeS$_2$ using neutron diffraction \cite{2001JNS...293.169} studies). According to Fig. \ref{GrGeGe} and Fig. \ref{GrSS} it seems that the decrease of the proportion
of homopolar bonds slows down for the lowest cooling rates and tends towards a limit of respectively 1.9$\%$ for the Ge atoms and 1.2$\%$ 
for the S atoms. These limiting values are small but nonzero and therefore our simulation results seem to confirm the existence
of homopolar bonds in experimental glassy GeS$_2$.\\
The simulation gives access to the positions of the atoms, therefore we can also obtain information on the connectivity of the network.
In our approach we focused on the ratio between edge and corner sharing tetrahedra and the evolution of the different local environments 
of the Ge (Tab. \ref{EnvGe}) and S atoms (Tab. \ref{EnvS}).\\
\begin{table}[ht]
\begin{center}
\begin{tabular}{ccccc}
\hline
\hline
\multirow{2}{2cm}{$\gamma$ (10$^{14}$K/s)} & \multicolumn{3}{c}{Proportion of Ge atoms[\%]} \\
 	&  {\bf Ge}(S$_4$) & {\bf Ge}(S$_3$) & {\bf Ge}(GeS$_3$) \\
\hline
\hline
 3   & 93.0 $\pm$2.0 & 2.3 $\pm$1.8 & 3.7 $\pm$2.1 \\
0.6  & 95.0 $\pm$1.6 & 1.7 $\pm$1.1 & 1.9 $\pm$2.0 \\
0.3  & 96.2 $\pm$1.8 & 0.3 $\pm$0.3 & 2.8 $\pm$2.0 \\
0.12 & 96.4 $\pm$1.8 & 1.7 $\pm$1.8 & 1.9 $\pm$2.0 \\
0.09 & 97.6 $\pm$1.7 & 0.1 $\pm$0.1 & 1.8 $\pm$1.0 \\
0.06 & 98.1 $\pm$1.0 & 0.0 $\pm$0.0 & 1.9 $\pm$1.1  \\
\hline
\end{tabular}
\caption{\label{EnvGe}Evolution of the local structural environment of Ge atoms as a function of the cooling rate $\gamma$}
\end{center}
\end{table}
 \begin{table}[ht]
\begin{center}
\begin{tabular}{cccccc}
\hline
\hline
\multirow{2}{2cm}{$\gamma$ (10$^{14}$K/s)} & \multicolumn{5}{c}{Proportion of S atoms[\%]} \\
 	&  {\bf S}(Ge$_2$) & {\bf S}(Ge) & {\bf S}(Ge$_3$)  & {\bf S}(GeS) & {\bf S}(Ge$_2$S) \\
\hline
\hline
 3   & 67.9 $\pm$2.2 & 14.3 $\pm$1.5 & 12.9 $\pm$1.8 & 3.1 $\pm$1.3 & 1.8 $\pm$0.9 \\
0.6  & 71.0 $\pm$4.1 & 12.9 $\pm$2.0 & 12.6 $\pm$1.8 & 2.1 $\pm$0.8 & 0.4 $\pm$0.3 \\
0.3  & 75.5 $\pm$2.6 & 11.4 $\pm$1.0 & 10.7 $\pm$1.2 & 2.0 $\pm$1.3 & 0.4 $\pm$0.6 \\
0.12 & 76.1 $\pm$2.7 & 10.9 $\pm$1.2 & 10.5 $\pm$1.3 & 1.7 $\pm$0.8 & 0.6 $\pm$0.6 \\
0.09 & 78.2 $\pm$3.3 & 10.4 $\pm$1.6 & 9.9 $\pm$1.4  & 1.1 $\pm$1.1 & 0.4 $\pm$0.3 \\
0.06 & 78.5 $\pm$1.0 & 10.4 $\pm$0.4 & 9.8 $\pm$1.0  & 0.8 $\pm$0.3 & 0.4 $\pm$0.3 \\
\hline
\end{tabular}
\caption{\label{EnvS}Evolution of the local structural environment of S atoms as a function of the cooling rate $\gamma$}
\end{center}
\end{table}
\noindent\\First it is worth noting that the proportion of edge and corner-sharing links 
is almost a constant independent of the cooling rate: 84$\%\pm$1.8 of corner-sharing and 16$\%\pm$1.8 of 
edge-sharing bounds, values in good agreement with experimental data \cite{1986PhRvB..33.5421B}. This is in contrast with the evolution of the proportions of Germanium (Tab.\ref{EnvGe}) and Sulfur (Tab.\ref{EnvS}) in their standard environment (respectively a four fold S coordination for Ge and a two fold Ge coordination for S) which appreciably increase with decreasing cooling rate.
The second point concerns, as expected, the decrease of the chemical disorder with decreasing cooling rate. 
Indeed the proportion of under-coordinated Ge atoms (2.3$\%$ for the fastest cooling rate) disappears for the slowest cooling rate. And the proportions of non-bridging S atoms and over-coordinated S atoms, respectively 14.25$\%$ and 12.85$\%$ decrease to 10.4$\%$ and 9.8$\%$.\\
These results indicate that the cooling rate has an impact on the structure of the glass. Nevertheless while certain
types of structural ``defects'' disappear at low rate, others survive and can therefore be considered as inherent of the glassy 
structure.  

\begin{figure}[h!]
\begin{center}
\includegraphics*[width=17cm, keepaspectratio=true, draft=false]{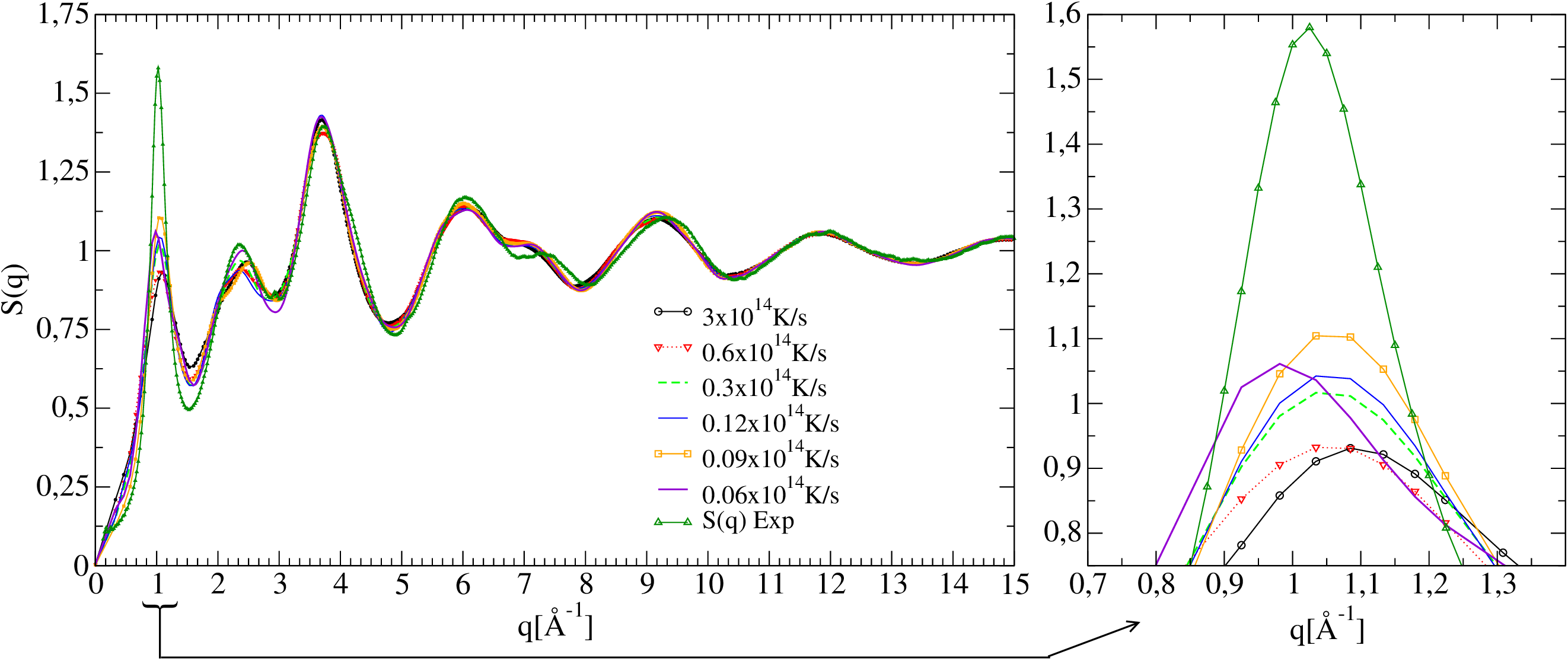}
\end{center}
\caption{\label{Sq}Total simulated static neutron structure factors. Enlargement corresponds to the FSDP part of the S(q).}
\end{figure}

\begin{figure}[h!]
\begin{center}
\includegraphics*[width=11cm, keepaspectratio=true, draft=false]{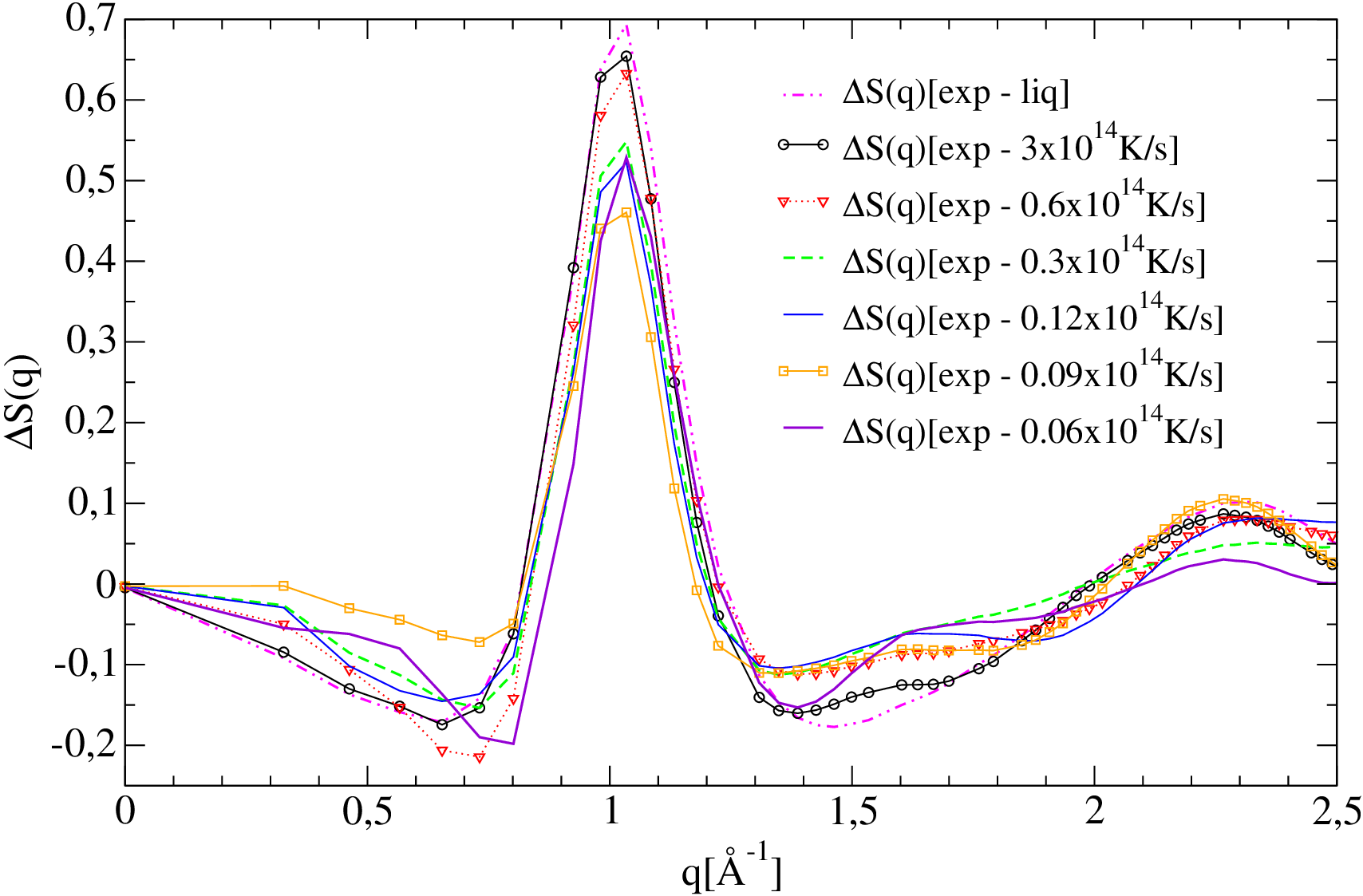}
\end{center}
\caption{\label{dsq} $\Delta$S(q) for the IRO, between experiment and liquid GeS$_2$, and between experiment and glassy GeS$_2$ for the different cooling rates.}
\end{figure}

\subsubsection{Neutron static structure factor}

\noindent\\\\An alternative way to analyze the structure is to compute the static neutron structure factor $S(q)$ which can be directly 
compared to neutron scattering experiments:\\
\begin{equation}
\label{sqeq}
S(q)=\frac{1}{N} \sum_{j,k} b_j\,b_k \left< e^{iq[r_j-r_k]} \right>
\end{equation}
where N is the number of atoms, and $b_j$ is the neutron scattering factor for atom $j$.\\
As for the radial pair distribution functions, we averaged for each cooling rate the total neutron structure factors 
of the five samples, permitting thus the comparison of the evolution of the average structure factors as a function 
of the cooling rate (Fig. \ref{Sq}).\\
First we note the accurate description of glassy GeS$_2$ reflected in the good agreement between the simulated curves and 
the experimental one, and this even for the highest cooling rates.
Although the simulated and experimental curves present differences 
in the range 0-2.5\AA$^{-1}$ it has already been shown \cite{2004PhRvB..70r4210B,2004PhRvB..69f4201B,blaineau-2006-} 
that the physical properties of the so-simulated glassy samples are in good agreement with experiment.
The First Sharp Diffraction Peak (FSDP), signature of the intermediate range order (IRO) in amorphous samples, 
appears at $\approx$1\AA$^{-1}$ and is, as expected in such simulations, slightly underestimated \cite{2003PhRvB..67i4204B}.
Size effects can be considered as an explanation, nevertheless
we note that a decrease of the cooling rate globally improves the calculated structure factor and in particular 
the FSDP (Fig. \ref{Sq}), which highlights that the IRO is also cooling rate dependent.
However there is no linear/regular evolution of S(q) with the cooling rate. 
To illustrate the improvement of S(q) in the FSDP region with the decrease of the cooling rate, we represent (Fig. \ref{dsq}) the 
difference, between 0 and 2.5\AA$^{-1}$, of the experimental total neutron structure factor and:\\
\begin{enumerate}[1) ]
\item the total neutron structure factor of the liquid state at 2000K:\\
 $\qquad \qquad S(q)_{exp} - S(q)_{liq} = \Delta S(q)_{exp-liq}$ 
\item the average total neutron structure factor of the simulated glass quenched down to 300K at the rate $\gamma$:\\
 $\qquad \qquad S(q)_{exp} - S(q)_{\gamma} = \Delta S(q)_{exp - \gamma}$
\end{enumerate}
The $\Delta S(q)_{exp-liq}$ is a reference representing the biggest variation between the experimental and simulated 
S(q) and should be compared to $\Delta S(q)_{exp - \gamma}$.
We see in Fig. \ref{dsq} that by decreasing the cooling rate, the difference $\Delta S(q)_{exp - \gamma}$ decreases. In 
particular it appears that for cooling rates higher than 0.3x10$^{14}$ K/s the 
differences $\Delta S(q)_{exp - \gamma}$ are very close to the difference $\Delta S(q)_{exp-liq}$ .
For the highest cooling rate (3x10$^{14}$ K/s) this difference is even quasi identical to $\Delta S(q)_{exp-liq}$ 
which indicates that there is no real change in the IRO between the fastest quenched glass and the liquid phase.
One can thus argue that the fastest cooled samples are too similar to the liquid and can therefore not be considered as glassy 
GeS$_2$ samples. This defines a limit for the maximum cooling rate usable in MD simulations in order to avoid interferences 
between the liquid and the glassy state. In our simulations this limit appears to be between 0.6x10$^{14}$ K/s and 
0.3x10$^{14}$ K/s. Nevertheless this limit is directly related to the method {\it i.e. ab-initio} MD simulations and 
indirectly related to our model {\it i.e.} the nature of the glass and the characteristics of the atomic pseudopotentials and
therefore the numerical value of this limit can not be straightforwardly extended to other simulated glassy systems.

\subsubsection{Atomic charges}

\noindent\\\\Even if atomic charges cannot actually be determined experimentally, relevant tools such as L\"owdin \cite{1950JChemPhys..18.365} 
or Mulliken \cite{1955JChemPhys..23.1833} population analysis can be used to compare different configurations with the 
{\it same} description.
In the present work the L\"owdin description has been chosen in order to compare the dependence of the atomic charges on 
the cooling rate.  It should however be mentioned that the non-self-consistent Harris functional is known to overestimate the 
charge transfers between the atoms.\\
The atomic charge {\it q} is calculated by the difference between the number of electrons of the neutral atom and the "real" 
number of electrons of the atom in the glass. We found no dependence of the L\"owdin charges on the cooling rate. Thus it
is necessary to correlate the charges with the evolution of the proportion of each atomic type in its local structural 
environment with the cooling rate (Tab. \ref{EnvGe} and \ref{EnvS}).\\
As expected the general polarity of the Ge-S bond is found with a charge transfer in an ordered Ge(S$_4$)$_{1/2}$ configuration 
of +0.94 for the Ge atoms and -0.46 for the S atoms. As it has already been shown in a previous work \cite{2004PhRvB..70r4210B} 
Ge charges are always positive and decrease with the number of neighbors whereas S charges are more variable with respect 
to the local environment: from strongly negative charges for non-bridging S atoms (-1.07) to almost neutral charges for 
3 fold Ge-coordinated S atoms. And even if the existence of positively charged S atoms in the environment 
{\large{$_{\tt{Ge}}^{\tt{Ge}}>$}}{\bf{S}}-S is confirmed in our present work, Tab. \ref{EnvS} shows that this kind of 
local structural environment disappears rapidly with decreasing cooling rate.
These structures obtained at high cooling rate may be explained by the results shown in Tab. \ref{ttble_q3}, representing the charge of an atom in a given local structural environment and the proportion of atoms in this environment in an equilibrium GeS$_2$ liquid at 2000K.
\begin{table}[ht]
\begin{center}
\begin{tabular}{ccc}
\hline
\hline
 Environment & {\it q} & Proportion \\
\hline
\hline
{\bf Ge}(S$_4$)   & +0.96$\pm$0.04 & 64.83$\%$ \\
{\bf Ge}(S$_3$)   & +1.02$\pm$0.06 & 27.41$\%$ \\
{\bf Ge}(GeS$_3$) & +0.75$\pm$0.1 & 1.14$\%$ \\
\\
Ge-{\bf S}-Ge  & -0.43$\pm$0.15 & 60.22$\%$ \\
Ge-{\bf S}    & -0.92$\pm$0.21 & 26.03$\%$ \\
{\large{$^{\tt{Ge}}_{\tt{Ge}}>$}}{\bf{S}}-Ge & -0.03$\pm$0.11 & 9.74$\%$ \\
Ge-{\bf{S}}-S & -0.1$\pm$0.18 & 2.07$\%$ \\
{\large{$_{\tt{Ge}}^{\tt{Ge}}>$}}{\bf{S}}-S & 0.32$\pm$0.2 & 1.60$\%$ \\
\hline
\end{tabular}
\caption{\label{ttble_q3}L\"owdin charges according to the local structural environments in an equilibrium liquid GeS$_2$ at 2000K}
\end{center}
\end{table}
Indeed it appears that the configuration of the liquid is quite similar, at least for the S atoms, to the fastest cooled glass configurations (Tab. \ref{EnvS}). This result indicates that the fastest cooled glasses are, in the literal sense, frozen liquids.
This confirms what we have already detected in the total neutron structure factor.\\
Positively and negatively charged zones inside the glass have been reported in our previous study \cite{2004PhRvB..70r4210B} 
for the highest cooling rate.
In order to see if the atomic charges measured for the lowest cooling rate confirm or reject the existence of such zones we looked at the short-range charge deviation $\Delta Q_{SR}$ of a particle $i$ defined by:
\begin{equation}
\label{eqdqsr}
\Delta Q_{SR}(i) = q(i) + \sum_{j=1}^{n(i)} \frac{q(j)}{n(j)}
\end{equation}
This allows to take into account the atomic charge $q(i)$ of a given particle $i$ as well as the charges on its $n(i)$ nearest neighbors (determined from the radial pair distribution function).
Whereas for a crystalline structure, in which no bond defects are present, this value is almost zero for all the particles, 
positive and negative values appear for glassy samples.\\
In Tab. \ref{dqsr} we have reported the evolution of the number of charged zones (a charged zone contains at least 2 nearest neighbors having the same sign for $\Delta Q_{SR}$) as a function of the cooling rate.\\
\begin{table}[ht]
\begin{center}
\begin{tabular}{cccc}
\hline
\hline
\multirow{2}{2cm}{$\gamma$ (10$^{14}$K/s)} & \multicolumn{2}{c}{Averaged number of charged zones} \\
 & $\Delta Q_{SR}$ $>$ +0.3 & $\Delta Q_{SR}$ $<$ -0.3 \\
\hline
\hline
 3   & 9.0 $\pm$1.0 & 20.8 $\pm$1.2 \\
0.6  & 9.0 $\pm$2.0 & 19.2 $\pm$0.8 \\
0.3  & 8.4 $\pm$1.6 & 17.2 $\pm$0.8 \\
0.12 & 8.6 $\pm$1.4 & 16.0 $\pm$1.0 \\
0.09 & 7.8 $\pm$1.8 & 15.8 $\pm$0.2 \\
0.06 & 9.2 $\pm$0.8 & 16.4 $\pm$0.6 \\
\hline
\end{tabular}
\caption{\label{dqsr}Evolution of the average number of charged zones as a function of the cooling rate $\gamma$}
\end{center}
\end{table}
\begin{table}[ht]
\begin{center}
\begin{tabular}{cccc}
\hline
\hline
\multirow{2}{2cm}{$\gamma$ (10$^{14}$K/s)} & \multicolumn{2}{c}{Averaged number of atoms per charged zones} \\
 & $\Delta Q_{SR}$ $>$ +0.3 & $\Delta Q_{SR}$ $<$ -0.3 \\
\hline
\hline
 3   & 5.7 $\pm$0.1 & 2.1 $\pm$0.0 \\
0.6  & 5.2 $\pm$0.4 & 2.1 $\pm$0.1 \\
0.3  & 5.2 $\pm$1.1 & 2.0 $\pm$0.0 \\
0.12 & 5.2 $\pm$0.2 & 2.1 $\pm$0.0 \\
0.09 & 5.2 $\pm$1.7 & 2.1 $\pm$0.0 \\
0.06 & 4.2 $\pm$0.2 & 2.1 $\pm$0.0 \\
\hline
\end{tabular}
\caption{\label{sdqsr}Evolution of the average number of atoms per charged zones as a function of the cooling rate $\gamma$}
\end{center}
\end{table}
The number of positively charged zones (particles with $\Delta Q_{SR}$ $>$ +0.3) is almost constant and equal to 9, 
with no dependence on the cooling rate, whereas the number of negatively charged zones (particles with $\Delta Q_{SR}$ $<$ -0.3) 
decreases slightly from $\sim$20 for the samples cooled at 3$\times$10$^{14}$ K/s to $\sim$16 for the samples cooled at a rate smaller 
than 0.3$\times$10$^{14}$ K/s (this is again coherent with the idea of a maximum cooling rate usable in MD simulations).
The negatively charged zones are principally made of Ge atoms coordinated to one or more non-bridging sulfur atoms, 
and for the fastest cooled glasses, of a few S-{\bf S}-Ge structures.
As already shown (Tab. \ref{EnvS}) these structures disappear with decreasing cooling rate.
 This observation correlated to the diminution of the proportion
of non-bridging sulfur atoms gives an explanation to the decrease of the number of negatively 
charged zones with the cooling rate for rates higher than 0.3$\times$10$^{14}$K/s.
The positively charged zones are exclusively made of Ge atoms linked to over-coordinated S atoms.\\
In addition we have reported the average number of atoms per charged zone as a function of the cooling rate (Tab. \ref{sdqsr}).
This shows that the size of the positively and the negatively charged zones (respectively with 5 and 2 atoms per zone) is 
independent of the cooling rate and therefore remains constant. It is worth noting that the global neutrality of the glass 
is always respected. Our results show that the existence of charged zones in glassy GeS$_2$ is not influenced by the 
variation of the cooling rate. They confirm thus those of our previous study, and show that extended charged zones (whose
manifestation has also been detected recently for other chalcogenide systems by Taraskin {\it et al.} \cite{PhysRevLett.97.055504})  
reflect the broken chemical order of the glass and are therefore inherent to the amorphous state.

\section{\label{conclu}Conclusion}

Through DFT based molecular dynamics simulations we have analyzed the effect of the cooling rate on some properties
 of glassy GeS$_2$. Influence of the cooling rate on the glass transition temperature as well as on the structural properties
 has been studied.\\
Due to a lack of statistics mainly due to computer time limitations (especially for the lowest cooling rates), the detailed 
variation of $T_g$ with the cooling rate could not be obtained. Nevertheless the extrapolation of our results to ``realistic'' 
cooling rates is in good agreement with the experimental glass transition temperature.\\
Analyzing the radial pair distribution functions and the local structural environments for each cooling rate, we find that 
the number of S and Ge homopolar bonds as well as the number of coordination defects decrease with the cooling rate.
Nevertheless the decrease of the homopolar bonds seems limited and therefore it is reasonable to think that this type of defect
is present in real glasses. However calculations at lower cooling rates should be done to confirm this observation.\\
The study of the simulated total neutron structure factor has confirmed the reliability of our model in the description 
of glassy GeS$_2$. We have analyzed the effect of the cooling rate on the intermediate range order whose signature is 
the first sharp diffraction peak. The simulated FSDP is closer to the experimental one for the slowest cooled glasses.
By comparing with the properties of the liquid state, we have shown that a maximum cooling rate should not be exceeded in the simulation 
in order to reproduce the IRO characteristic of the glassy phase.
The value of this maximum cooling rate will depend on the details of the model used to describe a given system.
The existence of a maximum cooling rate has also been supported by the analysis of the charges which has revealed that the 
electronic configuration of the fastest cooled glasses is close to the one obtained in liquid GeS$_2$. In addition the 
existence of positively and negatively charged regions in the amorphous state has been clearly confirmed even for the 
lowest cooled samples and seems therefore inherent to the glassy state. These regions will have an important impact 
on the properties of samples containing metallic ions as shown recently \cite{blaineau-2006-}.
 
\section*{Acknowledgments} 
The authors wish to thank S\'ebastien Blaineau for his help at the beginning of this work and 
Annie Pradel and Benoit Coasne for profitable discussions. Parts of the calculations have been performed at the ``Centre 
Informatique National de l'Enseignement Supérieur''(CINES) in Montpellier. 

%-----------------------------------------------------------------------------
\section*{References}

\bibliographystyle{unsrt}

\newpage

\section*{Figures}
%Figures

\end{document}